\documentclass[12pt]{article}
\usepackage{eqsection,epsfig,amsmath,cite}

\begin{document}

\newcommand{\<}{\langle}
\renewcommand{\>}{\rangle}
\def\r{\underline{\rho}}
\def\rh{\underline{\widehat{\rho}}}
\def\Q{{\cal Q}}
\def\Qh{\widehat{\cal Q}}
\def\d{\underline{\delta}}
\def\t{\widehat{t}}
\def\erf{{\rm erf}}

\title{On the nature of the low-temperature phase in\\
discontinuous mean-field spin glasses}

\author{{ Andrea Montanari}              \\
 {\small\it Laboratoire de Physique Th\'{e}orique de l'Ecole Normale
  Sup\'{e}rieure\footnote {UMR 8549, Unit{\'e}   Mixte de Recherche du 
  Centre National de la Recherche Scientifique et de 
  l' Ecole Normale Sup{\'e}rieure. } }
  \\[-0.2cm]
  {\small\it 24, rue Lhomond, 75231 Paris CEDEX 05, FRANCE}\\[-0.2cm]
  {\small E-mail: {\tt Andrea.Montanari@lpt.ens.fr}}
          \\[0.5cm]
  { Federico Ricci-Tersenghi}              \\
  {\small\it Dipartimento di Fisica and SMC and UdR1 of INFM}\\[-0.2cm]
  {\small\it Universit\`a di Roma "La Sapienza"}\\[-0.2cm]
  {\small\it Piazzale Aldo Moro 2, I-00185 Roma, ITALY}\\[-0.2cm]
  {\small E-mail: {\tt Federico.Ricci@roma1.infn.it}}
          \\[-0.1cm]
  {\protect\makebox[5in]{\quad}}  
  \\
}

\date{January 30, 2003}

\maketitle

\thispagestyle{empty}

\abstract{The low-temperature phase of discontinuous mean-field spin
glasses is generally described by a one-step replica symmetry breaking
(1RSB) Ansatz. The Gardner transition, i.e.\ a {\em
very}-low-temperature phase transition to a full replica symmetry
breaking (FRSB) phase, is often regarded as an inessential, and
somehow exotic phenomenon.  In this paper we show that the metastable
states which are relevant for the out-of-equilibrium dynamics of such
systems are {\em always} in a FRSB phase.  The only exceptions are (to
the best of our knowledge) the $p$-spin spherical model and the random
energy model (REM).  We also discuss the consequences of our results
for aging dynamics and for local search algorithms in hard
combinatorial problems.}

\clearpage

\section{Introduction and main results}
\label{IntroSection}

\begin{figure}
\centerline{\epsfig{figure=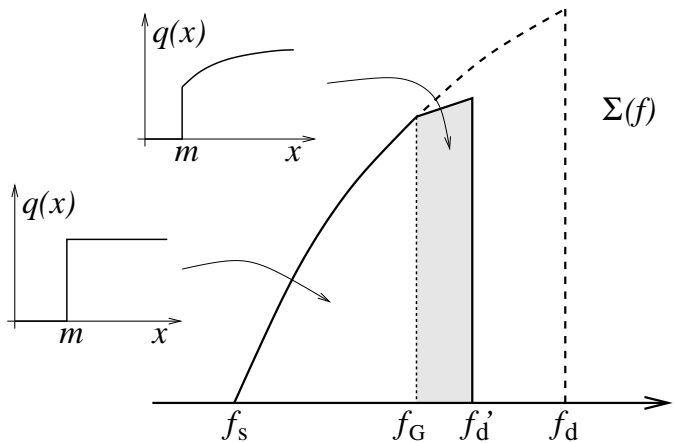,width=0.5\linewidth}}
\caption{The complexity curve for a generic discontinuous mean-field
spin glass. The dashed line is the 1RSB approximation, while the
continuous line is the exact result. The two coincide below $f_{\rm
G}$.  The gray region is FRSB. The shape of the Parisi order parameter
is shown in the insets.}
\label{GeneralPicture}
\end{figure}
During the last decade the $p$-spin spherical spin glass has been
thoroughly investigated both in its statical and in its dynamical
behavior~\cite{CrisantiSommers_Statics,CrisantiSommers_Dynamics,
CugliandoloKurchanPspin,DynamicsReview}.  In fact this model is
usually considered the prototypical example of discontinuous
mean-field spin glasses.  The latter, in turn, have been argued to be
crucial for understanding the structural glass
transition~\cite{KirkpatrickThirumalai1}.  In this paper we show that
the $p$-spin spherical model is indeed quite an exceptional case among
discontinuous mean-field models. While our results do not invalidate
the insight gained so far, they add some new important feature to the
general picture.

The out-of-equilibrium dynamics of the $p$-spin spherical models is
closely related to the structure of metastable states
\cite{KurchanMetastable}.  Below the dynamical temperature $T_{\rm
d}$, the Gibbs measure decomposes among an exponential number of
metastable states, whose free-energy densities lie between two values
$f_{\rm s}$ and $f_{\rm d}$.  The relevant quantity in this regime is
the complexity (or configurational entropy) $\Sigma_{T}(f)$. This is
defined in terms of the number ${\cal N}_T(f)$ of metastable states
having free energy density $f$,
\begin{eqnarray}
{\cal N}_T(f) \sim \exp\{N\Sigma_T(f)\}\, ,
\end{eqnarray}
$N$ being the number of degrees of freedom of the system.
$\Sigma_{T}(f)$ is strictly positive between $f_{\rm s}$ and $f_{\rm
d}$, with an absolute maximum at $f_{\rm d}$.  The free-energy density
$f_*(T)$ of the pure states which dominate the Gibbs measure, results
from the balance between energetic and entropic considerations.  While
the former would privilege the states at $f_{\rm s}$, the latter
favor the states at $f_{\rm d}$.

On the other hand, in a typical out-of-equilibrium set up, the system
is rapidly cooled below $T_{\rm d}$ from its high-temperature phase
\cite{CugliandoloKurchanPspin}.  In the thermodynamic limit, the
system never equilibrates and its behavior is dominated by the most
numerous metastable states, i.e.\ the ones at $f_{\rm d}$. This means
that, for any single-time observable ${\cal O}(t)$, the following
identity holds
\begin{eqnarray}
\lim_{t\to\infty}\lim_{N\to\infty}\<{\cal O}(t)\> = \lim_{N\to\infty}
\<{\cal O}\>_{f_{\rm d}}\, , \label{OneTimeQuantities}
\end{eqnarray}
where the average on the left-hand side is taken with respect to
thermal histories, while on the right-hand side it is taken with
respect to a constrained Gibbs measure. Outside the thermodynamic
limit, the system eventually equilibrates on an exponential time scale
$t_{\rm erg}(N) = \exp\{ O(N) \}$.

The complexity $\Sigma_T(f)$ can be calculated within a 1RSB scheme
\cite{MonassonMarginal}.  This calculation is known to be correct for
the $p$-spin spherical model \cite{CrisantiSommers_Statics}.  This
does not rule out the possibility of further replica-symmetry
breakings for more general cases: one would then expect a FRSB
calculation to be necessary.  The consequences of FRSB on the above
picture have not been investigated so far. However, it is usually
thought that FRSB would play a minor role.  The intuition, as far as
we can understand it, goes as follows.  As shown in 1984 by Elisabeth
Gardner \cite{Gardner} in the case of the $p$-spin Ising model,
discontinuous spin glasses may have a FRSB phase which overcomes the
1RSB phase at {\it very} low temperature.  The common wisdom
associates low temperature to low energy: if any FRSB effect is
present, it affects, at most, the lowest part of the complexity
spectrum. As a consequence, out-of-equilibrium dynamics is unaffected
by FRSB.  Here we show that this intuition is incorrect and that FRSB
plays an important role also for discontinuous mean-field glasses.

The correct scenario is sketched in Fig.~\ref{GeneralPicture}.  The
1RSB solution is stable with respect to FRSB up to some value $f_{\rm
G}$ of the free-energy density. At higher free energies, the 1RSB
solution becomes unstable and a FRSB calculation is required.  Quite
generally $f_{\rm G} < f_{\rm d}$ strictly.  We know just two
exceptions to this rule (both are somehow degenerate cases): the
spherical model and the REM \cite{GrossMezard}.  On the contrary both
the situations $f_{\rm G} < f_{\rm s}$ and $f_{\rm G} > f_{\rm s}$ are
possible.  A thermodynamic FRSB phase transition occurs when $f_{\rm
G}$ crosses $f_*(T)$.  Above $f_{\rm G}$, the FRSB calculation will
give a new complexity curve, and, in particular a new threshold free
energy $f_{\rm d}'$.  Of course, out-of-equilibrium dynamics will be
dominated (in the sense of Eq.  (\ref{OneTimeQuantities})) by the
states at $f_{\rm d}'$.  Finally, also the nature of aging dynamics
will change. We expect relaxation to be characterized by an infinite
number of time sectors \cite{CugliandoloKurchanSK,
CugliandoloKurchanWeak} (instead of just
two), although two of them will be most relevant (namely the first and
the last one).

In the next Sections we shall illustrate the general scenario with two
examples which are, at the same time, simple and representative.  In
Sec.~\ref{InfiniteConnSection} we reconsider the fully-connected Ising
$p$-spin model. The simplicity of this example allows us to study the
problem at finite temperatures. In Sec.~\ref{FiniteConnSection} we
turn to finite connectivity models at zero temperature. In this case
we can compare our results with numerical simulations.

\section{Infinite connectivity}
\label{InfiniteConnSection}

The fully-connected $p$-spin model is defined by the Hamiltonian
\begin{equation}
\mathcal{H}(\sigma) = -\sum_{(i_1 \dots i_p)}
J_{i_1 i_2 \ldots i_p} \sigma_{i_1} \sigma_{i_2} \ldots \sigma_{i_p}
\quad ,
\end{equation}
where the $N$ variables $\sigma_i=\pm1$ are Ising spins and the
$J_{i_1 i_2 \ldots i_p}$ are quenched random interactions extracted
from a Gaussian distribution with zero mean and variance $p!/(2
N^{p-1})$.  For illustration we shall often consider the $p=3$ case,
although our results are qualitatively valid for any finite $p>2$.

Using the standard replica formalism with a 1RSB Ansatz, one obtains
the action \cite{Gardner}
\begin{equation}
\phi(\beta,m) = -\frac{\beta}{4}\left[1+(p-1)(1-m)q^p-p\,q^{p-1}\right]
-\frac{1}{\beta}\log2 -\frac{1}{\beta m} \log \int \!\mathcal{D}z\,
\cosh^m\left(\beta\lambda\, z\right) \; ,
\label{eq:action}
\end{equation}
where $\mathcal{D}z \equiv e^{-z^2/2}\, dz/\sqrt{2\pi}$, $\lambda
\equiv\sqrt{\frac{p}{2}\,q^{p-1}}$ and $q$ is determined by the saddle
point equation
\begin{equation}
q = \frac{\int\! \mathcal{D}z\, \cosh^m(\beta\lambda z)
\tanh^2(\beta\lambda z)}{\int \!\mathcal{D}z \,\cosh^m(\beta\lambda
z)} \quad .
\label{eq:spe}
\end{equation}
From the action (\ref{eq:action}) we have the usual parametric
representation \cite{MonassonMarginal} of the complexity $\Sigma_T(f)$
at temperature $T=1/\beta$:
\begin{eqnarray}
\Sigma = \beta m^2 \partial_m \phi(\beta,m)\; , \qquad
f = \partial_m[m\, \phi(\beta,m)]\; .\label{Legendre}
\end{eqnarray} 
The outcome of this formulae agrees with the general picture sketched
in Fig.~\ref{GeneralPicture}, that is $\Sigma_T(f)$ is positive for $f
\in [f_{\rm s},f_{\rm d}]$, which corresponds to $m \in [m_{\rm
s},m_{\rm d}]$.  Both the extrema of this range depend on the
temperature.  The internal energy of metastable states can be
calculated using the identity
\begin{eqnarray}
e(\beta,m) = -\frac{1}{N}\sum_{(i_1, \dots , i_p)}\overline{
J_{i_1 i_2 \ldots i_p} \< \sigma_{i_1} \sigma_{i_2} \ldots
\sigma_{i_p}\>} = -\frac{\beta}{2}\Big[1-(1-m)\,q^p\Big]\; ,
\end{eqnarray}
where $q$ satisfies Eq.~(\ref{eq:spe}).

\begin{figure}
\centerline{\epsfig{figure=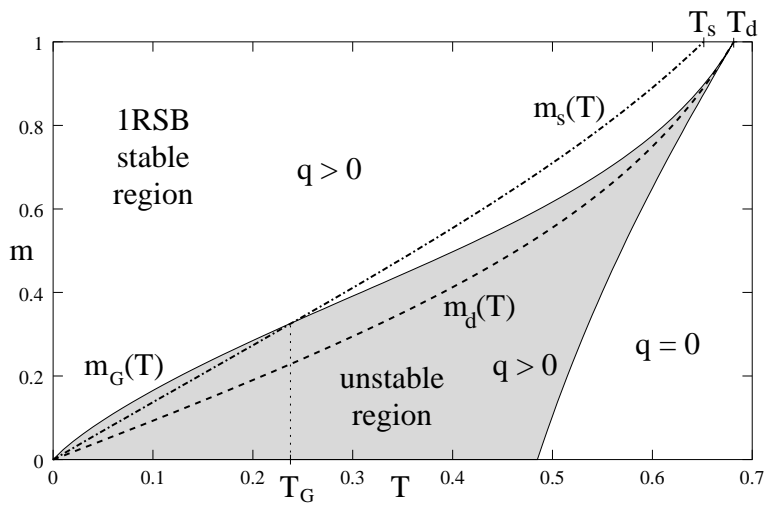,width=0.6\linewidth}}
\caption{3-spin fully-connected Ising model: the Parisi parameter $m$
for thermodynamic states (dotted-dashed curve) and for threshold
states (dashed curve) as a function of the temperature.  In the shaded
region, 1RSB solutions are unstable with respect to further replica
symmetry breakings.}
\label{fig:piano}
\end{figure}

The thermodynamics of the model at temperature $T$ (within the 1RSB
approximation) is obtained maximizing $\phi(T,m)$ with respect to
$m\in[0,1]$.  The resulting $m_{\rm s}(T)$ is shown in
Fig.~\ref{fig:piano} (dotted-dashed curve): it becomes smaller than 1
at the critical temperature $T_{\rm s}$ and it vanishes for $T\to 0$
as $m_{\rm s} \sim \mu_{\rm s} T$.  Threshold states, those maximizing
the complexity $\Sigma$, are obtained fixing $m=m_{\rm d}(T)$, which
is shown in Fig.~\ref{fig:piano} with a dashed curve: $m_{\rm d}(T)$
becomes smaller than 1 at the dynamical critical point $T_{\rm d}$ and
vanishes as $m_{\rm d} \sim \mu_{\rm d} T$.

The stability of the 1RSB solution with respect to a second step (and
eventually infinite steps) of replica symmetry breaking can be
evaluated following Ref.~\cite{Gardner}.  For any given $T$ and $m$
the 1RSB solution is stable provided
\begin{equation}
\frac{2}{p(p-1)\beta^2q^{p-2}} > \frac{\int \!\mathcal{D}z\,
\cosh^{m-4}(\beta\lambda z)}{\int
\mathcal{D}z \cosh^m(\beta\lambda z)}
\quad .
\end{equation}
In Fig.~\ref{fig:piano} the region where replica symmetry should be
broken more than once has been shaded.  We call $m_{\rm G}(T)$ its
boundary.  In Ref.~\cite{Gardner} Elizabeth Gardner calculated the
critical point $T_{\rm G}$ where thermodynamic states are no longer
1RSB (this point was called $T_2$ in Ref.~\cite{Gardner}).  Still more
interesting is the fact that {\em threshold states are always in the
unstable region}.

The physical scenario already described in Sec.~\ref{IntroSection}
(see Fig.~\ref{GeneralPicture}) is confirmed for any temperature below
$T_{\rm d}$.  In general we find that $f_{\rm G}$ is strictly less
than $f_{\rm d}$, while both the inequalities $f_{\rm G} < f_{\rm s}$
and $f_{\rm G} > f_{\rm s}$ are possible, depending on the
temperature.  For $f < f_{\rm G}$ the 1RSB solution is correct and the
Parisi order parameter $q(x)$ has a single step at $x=m$ (see lower
inset in Fig.~\ref{GeneralPicture} and please remind that in the
absence of an external field $q_0=0$).  For $f > f_{\rm G}$, the FRSB
calculation will give an order parameter $q(x)$ with a continuous
non-trivial part above the step (see upper inset in
Fig.~\ref{GeneralPicture}), a new complexity curve and, in particular,
a new threshold free energy $f_{\rm d}'$.  The geometrical structure
of metastable states above $f_{\rm G}$ is therefore the following.
There is an exponential number of {\it families} of states, each one
having a FRSB structure inside.  The typical overlap between two
families is 0, while the minimum overlap between states of the same
family is given by the step size in the function $q(x)$.

Let us suggest a natural generalization of the above method, for
computing $\Sigma(f)$ above $f_{\rm G}$.  Given the free-energy
functional \cite{SpinGlass} $\Phi_{\rm FRSB}[T,q(x)]$, one can
parameterize $q(x)$ by the jump location $m$ and by the continuous
function $r(x)$ for $x\in[m,1]$.  Then one can define
$\phi(T,m)=\max_{r(x)}\Phi_{\rm FRSB}[T,q(x)]$, by maximizing the
free-energy functional for fixed $T$ and $m$\footnote{This means that
one has to maximize $\Phi_{\rm FRSB}[T,q(x)]$ among all the order
parameters $q(x)$ such that $q(x)=0$ for $x<m$. Let us call $m_{\rm
s}^{\rm FRSB}$ the position of the discontinuity in the statical FRSB
solution $q_{\rm s}(x)$.  If $m<m_{\rm s}^{\rm FRSB}$, there is one
trivial solution to this maximization problem: $q(x)=q_{\rm s}(x)$. We
expect a non-trivial secondary maximum to exist.  On such a maximum
the discontinuity in $q(x)$ is located at $m$.  This expectation is
indeed confirmed by our 2RSB calculation.}.  Finally the complexity is
obtained by taking the Legendre transform of $\phi(T,m)$ as in
Eq. (\ref{Legendre}).  The complete calculation can be done using
e.g.\ the numerical method of Ref.~\cite{CrisantiRizzo}.  Here we show
the result of a 2RSB approximated calculation at zero temperature.

\begin{figure}
\centerline{\epsfig{figure=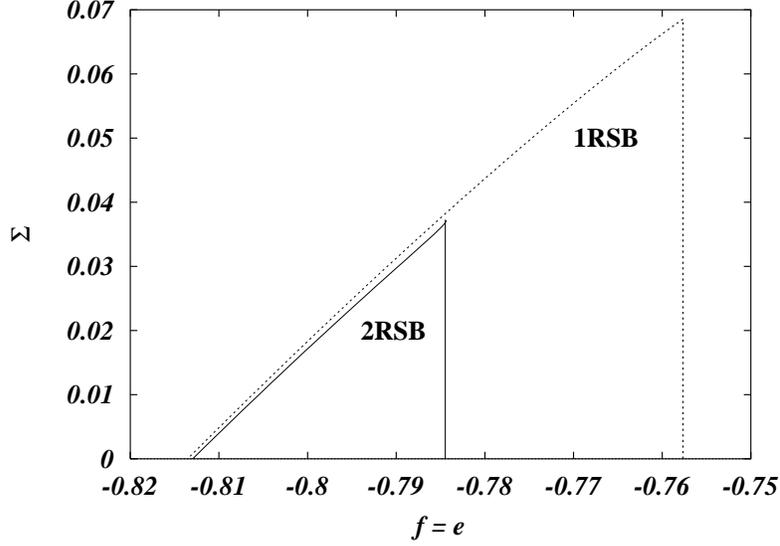,width=0.6\linewidth}}
\caption{$T=0$ complexity as a function of the (free) energy density
for the 3-spin fully-connected Ising model, obtained with 1RSB and
2RSB approximations.}
\label{fig:sigma2}
\end{figure}

In the $T \to 0$ limit, the 1RSB free energy is simply given by
\begin{equation}
\phi_{T=0}^{\rm 1RSB}(\mu) = -\frac{\mu}{4}-\frac1\mu \ln\left[
1+\erf\left(\frac{\sqrt{p}}{2}\mu\right)\right] \quad ,
\end{equation}
where $\mu = \lim_{\beta \to \infty} \beta m$ and $\erf(x) \equiv
2\int_0^x \! dt\, e^{-t^2}/\sqrt{\pi}$.  The resulting complexity
$\Sigma(e)$ is shown in Fig.~\ref{fig:sigma2} (dashed curve).

\begin{figure}
\centerline{\epsfig{figure=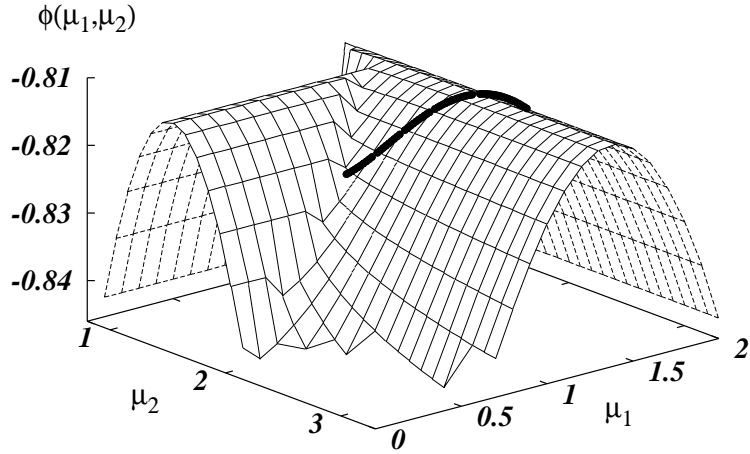,width=0.65\linewidth}}
\caption{Replicated free-energy for the 3-spin fully-connected Ising
model at $T=0$ as a function of the breaking parameters. The bold line
is the result of a maximization over $\mu_2$ at fixed $\mu_1$.  It
ends on the border of the region with $q=0$ (left paraboloid).  Notice
that the trivial maximum in this region has to be discarded when
computing the complexity, since it has $\partial_{\mu_1}
\phi(\mu_1,\mu_2)=0$.}
\label{fig:phi2}
\end{figure}

The 2RSB free energy $\phi_{T=0}^{\rm 2RSB}(\mu_1,\mu_2)$ depends on
two numbers which parameterize, as in the 1RSB case, the zero
temperature limit of the Parisi breaking parameters.  The three
overlaps behaves as follows: $q_0=0$, $q_1=q$ (with $0<q<1$ strictly)
and $q_2\simeq 1-\omega\, T$.  The saddle point equation for $q$
reads
\begin{eqnarray}
q &=& \frac{\int {\cal D}z\; I^{\nu-2}_+(z,q,\mu_2) \;
I^2_-(z,q,\mu_2)}{\int {\cal D}z\; I^\nu_+(z,q,\mu_2)}
\quad ,\label{eq:spe2}\\
I_\pm(z,q,\mu_2) &=& \frac{e^{\mu_2 \lambda z}}{2} \left[ 1 + \erf
\left(\frac{\eta \mu_2}{2} + \frac{\lambda z}{\eta} \right) \right]
\pm \frac{e^{-\mu_2 \lambda z}}{2} \left[ 1 + \erf \left(\frac{\eta
\mu_2}{2} - \frac{\lambda z}{\eta} \right) \right] \; ,
\end{eqnarray}
where $\nu \equiv \mu_1/\mu_2$, $\lambda \equiv \sqrt{\frac{p}{2}\,
q^{p-1}}$ and $\eta \equiv \sqrt{p(1-q^{p-1})}$.  At the saddle point,
the expressions for the action and the energy simplify to
\begin{eqnarray}
\phi(\mu_1,\mu_2) &=& -\frac{\mu_2}{4}\Big[ 1 + (p-1)(1-\nu)q^p -p\,
q^{p-1}\Big] - \frac{1}{\mu_1} \ln \int\!{\cal D}z \;
I_+^\nu(z,q,\mu_2) \; , \\
e(\mu_1,\mu_2) &=& -\frac12\Big[p\, \omega(q,\mu_2) + \mu_2 - (\mu_2 -
\mu_1) q^p\Big] \; ,\\
\omega(q,\mu_2) &=& \frac{2}{\sqrt{\pi p}} e^{-\frac14 \nu^2 \mu_2^2}
\frac{\int {\cal D}z\; I^{\nu-1}_+(z\sqrt{1-q^{p-1}},q,\mu_2)}{\int
{\cal D}z\; I^\nu_+(z,q,\mu_2)} \; ,
\end{eqnarray}
where, as usual, $q$ satisfies the saddle point equation
(\ref{eq:spe2}).  In the interesting region of parameters the shape of
$\phi(\mu_1,\mu_2)$ is that of 2 com-penetrating paraboloids (see
Fig.~\ref{fig:phi2}): On the left paraboloid we have that $q=0$ and we
recover the 1RSB solution with $\mu=\mu_2$, while on the right
paraboloid $q$ takes non-trivial values and we find there the absolute
maximum of $\phi(\mu_1,\mu_2)$, corresponding to the ground state
energy.  Please note that, on the right paraboloid, the most relevant
variable is the smaller breaking parameter $\mu_1$.

We proceed by maximizing $\phi(\mu_1,\mu_2)$ over $\mu_2$, and
Legendre-transforming the result with respect to $\mu_1$.  The result
is shown in Fig.~\ref{fig:sigma2} (continuous line).  It is worth
mentioning at least two important facts:
\begin{itemize}
\item The 2RSB threshold energy $e^{\rm 2RSB}_{\rm d}$ is much lower
than the 1RSB one $e^{\rm 1RSB}_{\rm d}$.
\item In the region between $e_s^{\rm 2RSB}$ and $e_d^{\rm 2RSB}$ the
complexity curve does not change significantly. In general we expect
$\Sigma(e)$ not to change below $e_{\rm G}$, but in this case $e_{\rm
G}<e^{\rm 2RSB}_{\rm s}$.
\end{itemize}

\begin{figure}
\centerline{\epsfig{figure=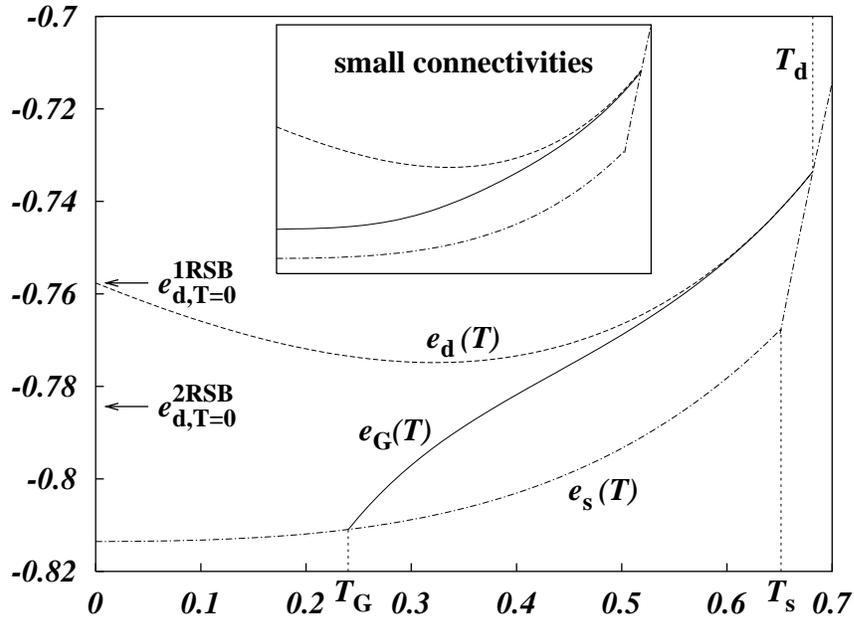,width=0.7\linewidth}}
\caption{Thermodynamic energy (dotted-dashed line) and threshold
energy (dashed line) as a function of the temperature within the 1RSB
approximation.  States above the full line are unstable with respect
to further replica symmetry breakings.  The arrows mark the threshold
energies calculated directly at $T=0$.  Main panel: 3-spin
fully-connected Ising model.  Inset: pictorial view for a 3-spin with
small connectivity.}
\label{fig:energie}
\end{figure}

How does this new physical scenario affect observable quantities, like
the internal energy?  In the main panel of Fig.~\ref{fig:energie} we
plot the thermodynamic energy $e_{\rm s}(T)$, the threshold states
energy $e_{\rm d}(T)$, and the internal energy $e_{\rm G}(T)$ at the
instability point, for the fully-connected 3-spin model.  The
incorrectness of the 1RSB Ansatz for threshold states is clear from
the unphysical behavior of $e_{\rm d}(T)$, which is not a monotonously
increasing function of $T$: an annealing experiment with a very slow
cooling, following 1RSB threshold states, would produce an increase of
energy when temperature is decreased!

The energy obtained with an extremely slow cooling, i.e.\ that of
threshold states, must lie below $e_{\rm d}(T)$ and above both $e_{\rm
s}(T)$ and $e_{\rm G}(T)$, and it must be a monotonously increasing
function of $T$.  We can estimate $e_{\rm d}(T=0)$, from the
zero-temperature 2RSB calculation above. The result is marked by an
arrow in Fig.~\ref{fig:energie}.  As already noticed, this value is
much lower than what predicted by an 1RSB calculations.  Such a
conclusion has positive consequences on the use of simulated annealing
and related techniques for the search of low energy solutions in hard
combinatorial problems.

However, the situation is not always like the one just described.  In
some cases, $e_{\rm G}(T)>e_{\rm s}(T)$ for all temperatures below
$T_{\rm d}$ (see the sketch in the inset of Fig.~\ref{fig:energie}).
This is what actually happens for $p$-spin models with small
connectivities, as will be shown in the next Section.

\section{Finite connectivity and numerical simulations}
\label{FiniteConnSection}

Finite connectivity mean-field models have been the object of intense
investigation in the last years. On one hand, they are thought to
share some properties of finite-dimensional models (namely each spin
interact with a finite number of neighbors).  On the other hand, they
are closely related to extremely hard combinatorial optimization
problems \cite{TCS}.  Nevertheless, up to now, the theoretical
investigations have been limited to the 1RSB level
\cite{MonassonRSB,MezardParisiBethe}.  Here we want to show how the
1RSB phase becomes unstable with respect to 2RSB fluctuations, and how
to compute the stability threshold.  We expect that, as usual, once
the 1RSB phase becomes unstable, FRSB has to be used for properly
describing the system.

For sake of simplicity, we present our calculation in a particularly
simple case, which allows a fully explicit derivation.  We shall
comment later on the generalizations of our results.  To be definite,
let us consider a fixed-connectivity Ising model with $p$-spin
interactions (hereafter $p>2$) with Hamiltonian
\begin{eqnarray}
{\cal H}(\sigma) = -\sum_{(i_1 \dots i_p)\in {\cal G}} J_{i_1 \dots
i_p} \sigma_{i_1} \dots \sigma_{i_p}\; .
\end{eqnarray}
In the above formula ${\cal G}$ is the hypergraph of interactions,
i.e. a set of $M$ among the $\binom{N}{p}$ possible $p$-uples of the
$N$ spins. Here we shall take ${\cal G}$ to have fixed connectivity:
each spin is supposed to participate to $(l+1)$ interaction terms.
Finally we shall consider the couplings $J_{i_1 \dots i_p}$ to take
the values $\pm 1$ with equal probability.

Under these hypothesis the 2RSB order parameter is a normalized
measure on a space of probability distributions and does not depend
upon the site of the sample \cite{FranzEtAlExact}.  The mean-field
equations are most conveniently written in terms of two such measures:
$\Q[\r]$ and $\Qh[\rh]$. At zero temperature they have the form
\begin{eqnarray}
\Q[\r] & = & \frac{1}{{\cal Z}}\int\! \prod_{\alpha=1}^{l}
d\Qh[\rh^{(\alpha)}]\,\, z[\{\rh^{(\alpha)}\}]^{\mu_1/\mu_2}\,
\delta\left[\r-\r^{(0)}[\{\rh^{(\alpha)}\}]\right]\,
,\label{Saddle1}\\
\Qh[\rh] & = & \int\!\prod_{i=1}^{p-1}d\Q[\r^{(i)}]\, \delta\left[ \rh
- \rh^{(0)}[\{\r^{(i)}\}]\right]\, ,\label{Saddle2}
\end{eqnarray}
where ${\cal Z}$ is a normalization constant and
$0<\mu_1\le\mu_2<\infty$ are the 2RSB parameters in the
$\beta\to\infty$ limit. The probability distribution $\rh$ is
supported over the integers $q$ such that $-1\le q \le +1$.  It is
therefore given in terms of three positive numbers: $\rh =
\{\widehat{\rho}_+,\widehat{\rho}_0,\widehat{\rho}_-\}$ (with
$\widehat{\rho}_+ +\widehat{\rho}_0 +\widehat{\rho}_- = 1$). The
distribution $\r$ is instead supported over the integers $-l\le q \le
l$. However, for our purposes, we can parametrize it using the three
numbers $\{\rho_+ \equiv\sum_{q=1}^l\rho_q,\rho_0,\rho_- =
\sum_{q=-l}^{-1}\rho_q\}$.  The ``functionals'' $\Q[\r]$ and
$\Qh[\rh]$ are therefore nothing but distributions over
two-dimensional simplexes.  Finally, the functions $\r^{(0)}[\dots]$
and $\rh^{(0)}[\dots]$ entering in
Eqs. (\ref{Saddle1})-(\ref{Saddle2}) are defined as follows:
\begin{eqnarray}
\rho^{(0)}_{+,0,-} & = & \frac{1}{z[\{\rh^{(\alpha)}\}]}
\mathop{\sum_{\{q_{\alpha}\}}}_{\sum q_{\alpha} >0, \, =0,\, <0}
\prod_{\alpha=1}^l\widehat{\rho}^{(\alpha)}_{q_i}\cdot
\exp\left\{-\mu_2\left[\sum_{\alpha=1}^l|q_i|-
|\sum_{\alpha=1}^l q_i|\right]\right\}\, ,
\label{SaddleRho1}\\
\nonumber\\
\widehat{\rho}^{(0)}_{q} & = &\left\{\begin{array}{ll}
\frac{1}{2}\left[\prod_{i=1}^{p-1} (\rho^{(i)}_+ + \rho^{(i)}_-)+
\prod_{i=1}^{p-1}(\rho^{(i)}_+ - \rho^{(i)}_-) \right] & \mbox{if
$q=+$}\, ,\\
& \\
1-\prod_{i=1}^{p-1} (\rho^{(i)}_++ \rho^{(i)}_-) & \mbox{if $q=0$}\,
,\\
& \\
\frac{1}{2}\left[\prod_{i=1}^{p-1} (\rho^{(i)}_+ + \rho^{(i)}_-)-
\prod_{i=1}^{p-1}(\rho^{(i)}_+ - \rho^{(i)}_-) \right] & \mbox{if
$q=-$}\, ,\\
\end{array}\right. 
\label{SaddleRho2}
\end{eqnarray}
and the constant $z[\{\rh^{(\alpha)}\}]$ is such that $\r^{(0)}$ is
normalized.

The authors of Ref.~\cite{FranzEtAlExact} considered the 1RSB solution
to this problem. This is nothing but the fixed point $(\r^*,\rh^{\,
*})$ of Eqs. (\ref{SaddleRho1}) and (\ref{SaddleRho2}):
\begin{eqnarray}
\r^* = \r^{(0)}[\rh^{\, *},\dots,\rh^{\, *}]\, ;\,\,\,\,\,\,\,
\rh^{\, *} = \rh^{(0)}[\r^*,\dots,\r^*]\, .\label{1RSBSolution}
\end{eqnarray}
The above equations have always a symmetric solution: $\rho^*_+ =
\rho^*_-$, $\widehat{\rho}^{\, *}_+ = \widehat{\rho}^{\, *}_-$, which
is the physical one.  The physical stability of the 1RSB phase
coincides with the stability of this solution under the iteration
(\ref{Saddle1})-(\ref{Saddle2}).  We must therefore consider how the
2RSB order parameters $\Q[\r]$, $\Qh[\rh]$ can reduce to the 1RSB
solution (\ref{1RSBSolution}):
\begin{itemize}
\item The first possibility is that $\Q[\r]$ and $\Qh[\rh]$
concentrate around $\r^*$, $\rh^{\, *}$:
\begin{eqnarray}
\Q[\r] \approx f(\r-\r^*)\, , \;\;\;\; \Qh[\rh] \approx
\widehat{f}(\rh-\rh^{\, *})\, ,
\label{FirstLimit}
\end{eqnarray}
where $f(\cdot)$ and $\widehat{f}(\cdot)$ are supported in a
neighborhood of $\underline{0}$. The two distributions on the
two-dimensional simplex $\Q$, $\Qh$ tend, in the 1RSB limit, to delta
functions in a particular point $(\r^*,\rh^{\, *})$ of the simplex.

The stability condition of the delta-function solution under
perturbations of the type (\ref{FirstLimit}) is easily derived. Define
the $2\times 2$ matrices $L$ and $\widehat{L}$ by linearizing
Eqs. (\ref{SaddleRho1}), (\ref{SaddleRho2}) around $(\r^*,\rh^{\,
*})$:
\begin{eqnarray}
L_{q,q'} = \left.\frac{\partial \rho^{(0)}_q}{\partial
\widehat{\rho}^{(1)}_{q'}} \right|_{\r^*}\, ,\;\;\;\;\;\;
\widehat{L}_{q,q'} = \left.\frac{\partial
\widehat{\rho}^{(0)}_q}{\partial \rho^{(1)}_{q'}}\right|_{\rh^{\,
*}}\, ,\;\;\;\;\;\;\;\;\;\;\;\; \mbox{for $q,q'\in\{+,-\}$}\, .
\end{eqnarray}
If we call $\lambda_{\rm MAX}$ the eigenvalue of their product
$L\cdot\widehat{L}$ having the maximum absolute value, we obtain the
stability condition
\begin{eqnarray}
l(p-1)\cdot |\lambda_{\rm MAX}| <1\, .\label{Criterion_1}
\end{eqnarray}
In fact the matrices $L$ and $\widehat{L}$ can be easily diagonalized
by using symmetry considerations: one of the two eigenvectors is
symmetric, and the other is antisymmetric under the exchange 
$+ \leftrightarrow -$. The
antisymmetric eigenvalue vanishes for $p>2$. The symmetric one can be
shown to verify always the stability condition (\ref{Criterion_1}).
This type of instability is therefore irrelevant for the problem under 
study.
\item The second possibility is that the functionals $\Q[\r]$ and
$\Qh[\rh]$ concentrate around delta-function distributions:
\begin{eqnarray}
\Q[\r] \approx \rho^*_+ f_+[\r-\d_+] + \rho^*_0\, f_0[\r-\d_0]+
 \rho^*_- f_-[\r-\d_-]\, ,
\end{eqnarray}
and analogously for $\Qh[\rh]$. In the above expression the
distributions $f_q[\cdot]$ are supported around $\underline{0}$, and
$\d_+$ (respectively $\d_0$, $\d_-$) is the distribution $\{1,0,0\}$
(respectively $\{0,1,0\}$, $\{0,0,1\}$). In other words the measures
$\Q[\r]$ and $\Qh[\rh]$ concentrate over the corners of the simplex,
with weights given by the 1RSB solution.

In order to derive the stability condition it is convenient to use
variables which are small near the corners of the simplex. One
possible choice is to rewrite the distributions $f_q[\cdot]$ as
functions of the variables
\begin{eqnarray}
\epsilon_{\pm} = \rho_{\pm} - \delta_{q,\pm}\, ,
\end{eqnarray}
and analogously for the distributions $\widehat{f}_q[\cdot]$.  Notice
that two variables are sufficient because of the normalization
constraint. Moreover the signs of $\epsilon_+$, $\epsilon_-$ are fixed
by the value of $q$.  It is now easy to linearize the equations for
$f_q[\cdot]$, $\widehat{f}_q[\cdot]$ in the limit $\epsilon_{\pm}\ll
1$. If we define $\<\epsilon_{\pm}\>_q$ (respectively
$\<\widehat{\epsilon}_{\pm}\>_{\widehat{q}}$), the average of
$\epsilon_{\pm}$ ($\widehat{\epsilon}_{\pm}$) with respect to
$f_q[\cdot]$ ($\widehat{f}_{\widehat{q}}[\cdot]$), we get the
equations
\begin{eqnarray}
\<\epsilon_{\sigma}\>_q \approx l\sum_{\widehat{q},\widehat{\sigma}}
T_{q\sigma,\widehat{q}\widehat{\sigma}}\,
\<\widehat{\epsilon}_{\widehat{\sigma}} \>_{\widehat{q}}\,
,\;\;\;\;\;\;\;\;\;\;\; \<\widehat{\epsilon}_{\widehat{\sigma}}
\>_{\widehat{q}}\approx (p-1)\sum_{q,\sigma}
\widehat{T}_{\widehat{q}\widehat{\sigma},q\sigma}\,\<\epsilon_{\sigma}\>_q\,
,
\label{Linearization}
\end{eqnarray}
where the sums over $q$ and $\widehat{q}$ run over $\{+,0,-\}$, while
the ones over $\sigma$ and $\widehat{\sigma}$ run over $\{+,-\}$.
Explicit formulae for the $6\times 6$ matrices $T$ and $\widehat{T}$
are reported in App.~\ref{AppendixMatrices}.  As before, we look for
the eigenvalue of the product $T\cdot \widehat{T}$ which has the
largest absolute value.  If we call $\omega_{\rm MAX}$ this
eigenvalue, the stability criterion is
\begin{eqnarray}
l(p-1)\cdot |\omega_{\rm MAX}|<1\, . \label{Criterion_2}
\end{eqnarray}
It turns out that $\omega_{\rm MAX}$ depends uniquely on the parameter
$\mu_2$, which can be identified as the 1RSB parameter.  This
criterion, unlike (\ref{Criterion_1}), has a non-trivial content which
we shall consider in the following.
\end{itemize}

The criterion (\ref{Criterion_2}) select a range of the 1RSB parameter
for which the 1RSB solution is stable. This range has the form $\mu>
\mu_{\rm G}$. Using the relation between $\mu$ and the energy $e$ of
metastable states \cite{MonassonMarginal}, see also the previous
Section, one can convert this condition as an energy condition of the
form $e<e_{\rm G}$. This confirms our general picture summarized in
Fig.~\ref{GeneralPicture}.

In Table~\ref{p3Table} we exemplify the general situation by
considering the case $p=3$. In the low connectivity regime $(l+1)\le
10$ we have $e_{\rm s} < e_{\rm G} < e_{\rm d}$ strictly: the ground
state is correctly described by 1RSB while high-lying metastable
states are unstable to FRSB. The expected temperature dependence of
$e_{\rm d}$ in this regime of connectivities, is sketched in the inset
of Fig. \ref{fig:energie}.  At higher connectivities the ground state
becomes unstable too: $e_{\rm G} < e_{\rm s} < e_{\rm d}$.  We know
that the last situation is verified in the infinite-connectivity
limit, cf. Sec.~\ref{InfiniteConnSection}.  Moreover, for even
connectivities, the instability point corresponds to the vanishing of
the 0-component of the 1RSB parameter $\widehat{\rho}^{\, *}_0$, and
thus the ground state become unstable when $\widehat{\rho}^{\,
*}_0(\mu_s)$ becomes positive.

\begin{table}
\centering\begin{tabular}{|l|l|l|l|l|l|l|l|}
\hline
$(l+1)$ & $\mu_{\rm d}$ & $-e_{\rm d}$ & $\mu_{\rm s}$ & $-e_{\rm s}$
& $\mu_{\rm G}$ & $-e_{\rm G}$ & $\widehat{\rho}^{\, *}_0(\mu_s)$\\
\hline
\hline
3  & 1.42832 & 0.958659 & $\infty$ & 1.0     & 1.70267  & 0.963594 & 0\\
4  & 0.915569 & 1.15267 & 1.4115   & 1.21771 & 1.09861  & 1.16667 & 0\\
5  & 0.716089 & 1.3254  & 1.09566  & 1.39492 & 0.958971 & 1.35728 & 0.0420615\\
6  & 0.613587 & 1.45936 & 0.901568 & 1.54414 & 0.804719 & 1.5     & 0\\
7  & 0.535146 & 1.59825 & 0.802528 & 1.68623 & 0.75739  & 1.66103 & 0.0427\\
8  & 0.49023  & 1.70826 & 0.717919 & 1.8092  & 0.677627 & 1.77828 & 0\\
9  & 0.445068 & 1.8279  & 0.663636 & 1.93191 & 0.652821 & 1.92257 & 0.0397177\\
10 & 0.41937  & 1.924   & 0.61494  & 2.03932 & 0.601444 & 2.02455 & 0\\
11 & 0.389004 & 2.03054 & 0.578677 & 2.14895 & 0.585751 & 2.15734 & 0.0366356\\
12 & 0.372092 & 2.11725 & 0.54668  & 2.2457  & 0.548851 & 2.2488  &0.00075104\\
13 & 0.34986  & 2.21409 & 0.51988  & 2.34567 & 0.537826 & 2.37296 & 0.0339348\\
14 & 0.337728 & 2.2939  & 0.497043 & 2.43451 & 0.509535 & 2.4566  &0.00375363\\
15 & 0.320558 & 2.38318 & 0.476086 & 2.52697 & 0.501257 & 2.5739  & 0.0316323\\
16 & 0.311329 & 2.45761 & 0.458822 & 2.60963 & 0.4786   & 2.65142 &0.00579464\\
17 & 0.297566 & 2.54079 & 0.441831 & 2.696   & 0.472092 & 2.7631  & 0.0296691\\
18 & 0.290241 & 2.61089 & 0.428229 & 2.77367 & 0.453371 & 2.83566 &0.00720516\\
19 & 0.278903 & 2.689   & 0.414083 & 2.855   & 0.44808  & 2.94261 & 0.0279824\\
20 & 0.272901 & 2.75551 & 0.40303  & 2.9285  & 0.432245 & 3.01104 &0.00820489\\
\hline
\end{tabular}
\caption{3-spin with fixed connectivity $(l+1)$ at $T=0$ within the
1RSB approximation: columns from 2 to 7, breaking parameter and energy
corresponding to threshold (d), thermodynamic (s) and marginally-stable (G)
states; column 8, 0-component of the 1RSB parameter on the states
dominating the thermodynamics.}
\label{p3Table}
\end{table}

For $(l+1)\le 10$, we are in the case depicted in the inset of
Fig.~\ref{fig:energie}, since the instability energy $e_{\rm G}$ lies
above the ground state energy $e_{\rm s}$.  In this situation there is
a lower bound on the energy reachable by simulated annealing (and
presumably by any local search algorithm running in a time polynomial
in $N$) which is strictly above the ground state energy, and the
problem of finding the ground state is hard and not approximable
\cite{Arora}.

\begin{figure}
\centerline{\epsfig{figure=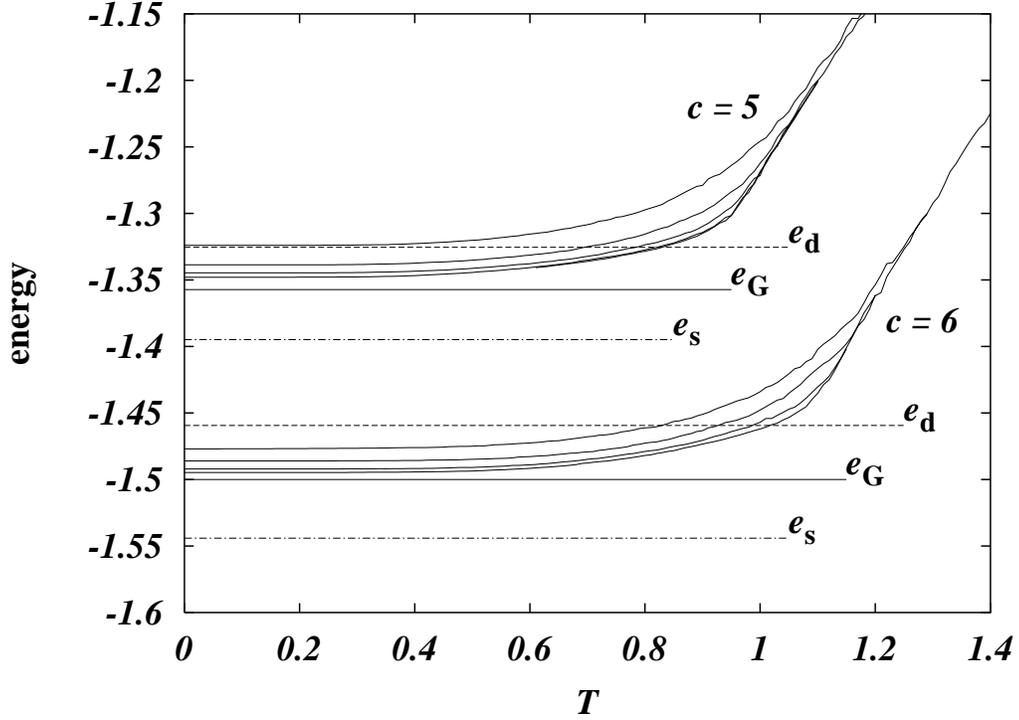,width=0.8\linewidth}}
\caption{3-spin Ising model with fixed connectivity $c$, energy
relaxation during slow simulated annealings.  The 4 different full
curves correspond to 4 cooling rates proportional to
1,$10^{-1}$,$10^{-2}$,$10^{-3}$.  The values for $e_{\rm d}$, $e_{\rm
G}$ and $e_{\rm s}$ are those reported in Table~\ref{p3Table}.}
\label{fig:cool}
\end{figure}

We have verified this prediction by running long simulated annealings
on a large instance of the 3-spin model with fixed connectivities 5
and 6.  The results are shown in Fig.~\ref{fig:cool}, different curves
corresponding to different cooling rates.  For both connectivities,
the energies reachable by simulated annealing are clearly below the
threshold energy $e_{\rm d}$ predicted with the 1RSB Ansatz.  Moreover
the extrapolations to infinitely slow cooling rates are perfectly
compatible with the energy $e_{\rm G}$, thus suggesting that the
complexity coming from the FRSB solution at $x>m$ should be  tiny.

\begin{figure}
\centerline{\epsfig{figure=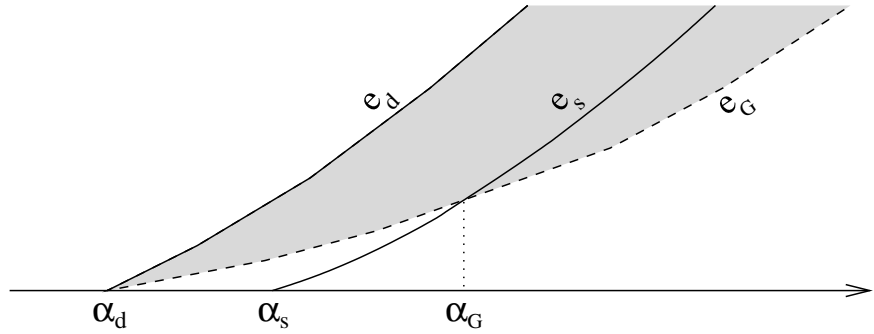,width=0.65\linewidth}}
\caption{The 1RSB zero-temperature phase diagram for spin models with
Poissonian connectivity.  The shaded area is an educated guess for
range of parameter in which FRSB is needed.}
\label{KsatFig}
\end{figure}

Any fixed-connectivity $p$-spin Ising model is therefore FRSB for what
concerns dynamic states.  Another interesting class of diluted
mean-field models consists of models defined on random hypergraphs
with Poissonian connectivity.  Among the others, this class includes
random 3-SAT \cite{3SAT,MarcGiorgioRiccardo,MarcRiccardo} and random
$p$-XORSAT \cite{XorSat,CoccoEtAlXorSat,MezardEtAlXorSat}.  The
zero-temperature phase diagram for these models in 1RSB approximation
is sketched in Fig.~\ref{KsatFig} (we recall that, for both these
models, energies are always positive defined).  Glassy metastable
states develop above the average connectivity $\alpha_{\rm d}$ and
have energy densities between $e_{\rm s}(\alpha)$ and $e_{\rm
d}(\alpha)$.  The ground state energy $e_{\rm s}(\alpha)$ becomes
positive at $\alpha_{\rm s}$ ($> \alpha_{\rm d}$).  A quite natural
conjecture for the energy above which FRSB sets in, $e_{\rm
G}(\alpha)$, is reported with a dashed line.  We have $e_{\rm s}<
e_{\rm G}<e_{\rm d}$ for $\alpha_{\rm d}<\alpha <\alpha_{\rm G}$, and
$e_{\rm G}< e_{\rm s}<e_{\rm d}$ for $\alpha>\alpha_G$.  The
calculation of the $e_{\rm G}(\alpha)$ curve will be presented in a
forthcoming publication \cite{Fluttuante}.

\section*{Acknowledgments}

It is a pleasure to thank Marc M\'ezard for his interest in our work.
We also acknowledge financial support from ESF programme SPHINX.

\appendix
\section{Formulae for the stability matrices}
\label{AppendixMatrices}

In this Appendix we give explicit formulae for the matrices $T$ and
$\widehat{T}$ implicitly defined by Eq. (\ref{Linearization}).  All
these formulae are expressed in terms of the 1RSB solution, cf. Eq.
(\ref{1RSBSolution}).  The form of the matrices is the following:
\begin{eqnarray}
T = \left[\begin{array}{cccccc}
t_1 & t_2 & 0 & t_3 & 0 & 0\\
0 & t_1 & 0 & 0 & 0 & 0\\
0 & 0 & t_4 & 0 & 0 & t_5\\
t_5 & 0 & 0 & t_4 & 0 & 0\\
0 & 0 & 0 & 0 & t_1 & 0\\
0 & 0 & t_3 & 0 & t_2 & t_1\\
\end{array}\right]\, ,\;\;\;\;\;\;\;\; 
\widehat{T}  = \left[\begin{array}{cccccc}
\t_1 & 0 & 0 & 0 & 0 & \t_1 \\
0 & \t_1 & 0 & 0 & \t_1 & 0 \\
0 & 0 & \t_2 & \t_2 & 0 & 0 \\
0 & 0 & \t_2 & \t_2 & 0 & 0 \\
0 & \t_1 & 0 & 0 & \t_1 & 0 \\
\t_1 & 0 & 0 & 0 & 0 & \t_1 \\
\end{array}\right]\, ,
\label{Matrices}
\end{eqnarray}
where we ordered the entries in Eq. (\ref{Linearization}) as follows:
$[\<\epsilon_+\>_+,\<\epsilon_-\>_+;\<\epsilon_+\>_0,\<\epsilon_-\>_0;
\<\epsilon_+\>_-,\<\epsilon_-\>_-]$.  The non-zero entries in the
above matrices are given below
\begin{align}
t_1 &= \frac{\Omega^{(0)}_1}{Z_+} &
t_2 &= -e^{-2\mu_2} \frac{\Omega^{(1)}_1}{Z_+} \quad\;
t_3 =  -e^{-2\mu_2} \frac{\Omega^{(1)}_0}{Z_+} &
t_4 &= \frac{\Omega^{(0)}_0}{Z_0} &
t_5 &= -e^{2\mu_2} \frac{\Omega^{(1)}_{-1}}{Z_0} \\
&& \t_1 &= \frac{1}{2}\, \frac{\rho_0^*(1-\rho_0^*)^{p-2}}
{1-(1-\rho_0^*)^{p-1}} &
\t_2 &= \, \frac{1}{2} &&
\end{align}
where we used the shorthands
\begin{eqnarray}
Z_{+,0,-} & \equiv & \mathop{\sum_{\{q_i\}}}_{\sum q_i >0, \, =0,\,
<0} \prod_{i=1}^{l-1}\widehat{\rho}_{q_i}^{\, *}\cdot
\exp\left\{-\mu_1\left[\sum_i|q_i|-|\sum_i q_i|\right]\right\}\, ,\\
\Omega^{(q)}_{\widehat{q}} & \equiv &
\widehat{\rho}^{\,*}_{\widehat{q}} \mathop{\sum_{\{q_i\}}}_{\sum
q_i=q} \prod_{i=1}^{l-2}\widehat{\rho}^{\, *}_{q_i}\cdot
\exp\left\{-\mu_1\left[|\widehat{q}|+\sum_i|q_i|-|\widehat{q}+ \sum_i
q_i|\right]\right\} = \nonumber\\
& = & \widehat{\rho}^{\,*}_{\widehat{q}}
\exp\Big[\mu_1\Big(|\widehat{q}+q|-|\widehat{q}|\Big)\Big]
\mathop{\sum_{\{q_i\}}}_{\sum q_i=q} \prod_{i=1}^{l-2}
\widehat{\rho}^{\,*}_{q_i}\: e^{-\mu_1 |q_i|} \; .
\end{eqnarray}
Let us finally notice that the $+/-$ symmetry can be exploited to
reduce the matrices (\ref{Matrices}) to $3\times 3$ matrices.

\end{document}